\documentstyle[12pt,aasms4]{article}

\def\ltsima{$\; \buildrel < \over \sim \;$}
\def\simlt{\lower.5ex\hbox{\ltsima}}

\input psfig.tex
\begin{document}
\onecolumn

\title{
Classifications of the Host Galaxies of Supernovae, Set III
}

\author{Sidney van den Bergh}
\affil{Dominion Astrophysical Observatory, Herzberg Institute of Astrophysics,
National Research Council, 5071 West Saanich Road, Victoria, British
Columbia, V9E 2E7, Canada (sidney.vandenbergh@nrc.gc.ca)
}

\centerline{and}

\author{Weidong Li and Alexei V. Filippenko}
\affil{
Department of Astronomy, 601 Campbell Hall, University of
California, Berkeley, CA 94720-3411 (wli@astro.berkeley.edu,
alex@astro.berkeley.edu)
}

\begin{abstract}

A homogeneous sample comprising host galaxies of 604 recent
supernovae, including 212 objects discovered primarily in 2003 and 2004, 
has been classified on the David Dunlap Observatory system.
Most SN 1991bg-like SNe Ia occur in E and E/Sa galaxies, whereas the
majority of SN 1991T-like SNe~Ia occur in intermediate-type galaxies.
This difference is significant at the 99.9\% level. As expected,
all types of SNe~II are rare in early-type galaxies, whereas
normal SNe~Ia occur in all Hubble types. This difference is
significant at the 99.99\% level. A small number of SNe~II in
E galaxies might be due to galaxy classification errors, or to a
small young-population component in these mainly old objects.
No significant difference is found between the distributions
over Hubble type of SNe~Ibc and SNe~II. This confirms that
both of these types of objects have similar (massive)
progenitors. The present data show that, in order to understand
the dependence of supernova type on host-galaxy population, it
is more important to obtain accurate morphological
classifications than it is to increase the size of the data
sample.

\end{abstract}

 
\section{The Lick Observatory Supernova Search}

     The present paper represents a continuation of the investigations by van
den Bergh, Li, \& Filippenko (2002, 2003; hereafter Papers I and II), in which
we studied the morphologies of the host (parent) galaxies of supernovae (SNe)
that were discovered (or independently rediscovered) during the course of the
Lick Observatory Supernova Search (LOSS) with the 0.76~m Katzman Automatic
Imaging Telescope (KAIT).\footnote{Although most of the SNe were 
discovered by LOSS itself, some were first discovered and reported by other
observers. Nevertheless, their host galaxies were being monitored with LOSS
at the time of discovery, and LOSS independently recognized the new objects 
as SNe; thus, they are included in the complete sample.}
This is the first step in the LOSS-based calculation
of rates of various types of SNe, currently being conducted by Leaman, Li, \&
Filippenko (2004). 

   LOSS, which started in March 1997 (Treffers et al. 1997), has been described
by Li et al. (2000), Filippenko et al. (2001), and Filippenko (2003,
2005). During the interval late-October 2000 through mid-October 2003, it was
expanded to the Lick Observatory and Tenagra Observatory Supernova Searches
(LOTOSS; Schwartz et al. 2000), but thereafter it reverted back to simply LOSS
(Filippenko et al. 2003), using KAIT alone, without the assistance of Tenagra
Observatory.

    KAIT is a fully robotic instrument whose control system checks the weather,
opens the dome, points to the desired objects, acquires guide stars (in the
case of long exposures), exposes, stores the data, and manipulates the data
automatically, all without human intervention. We reach a limit of $\sim$19 mag
($4\sigma$) in 25-s unfiltered, unguided exposures (used in the supernova
search), while 5-min guided exposures yield $R \approx 20$ mag.  Besides
conducting a supernova search, KAIT acquires well-sampled, long-term light
curves of SNe and other variable or ephemeral objects --- projects that are
difficult to conduct at other observatories having a large number of users with
different interests.

   Special emphasis is placed on finding SNe well before maximum
brightness. Although the original LOSS sample had only about 5000 galaxies, in
the year 2000 we increased the sample to $\sim$14,000 galaxies (most with
recession speed $cz \simlt 10,000$ km s$^{-1}$), separated into three subsets (observing
baselines of 2 days for about 100 galaxies, 3--6 days for $\sim$3000 galaxies,
and 7--14 days for $\sim$11,000 galaxies). In early June 2004, we
decreased the sample to 7500 galaxies, in order to have a shorter baseline and be
better able to determine the explosion date accurately. Specifically, we
adopted this last strategy to find SNe~Ia for an extensive study of their
ultraviolet properties with the {\it Hubble Space Telescope (HST)} ---
GO--10182 (P.I.: Filippenko).

    We are able to observe $\sim$1000 galaxies per night in unfiltered
mode. Our software automatically subtracts new images from old ones (after
registering, scaling to account for clouds, convolving to match the
point-spread-functions, etc.), and identifies SN candidates which are
subsequently examined and reported to the Central Bureau for Astronomical
Telegrams by numerous research assistants (mostly undergraduate students) 
in our group at the University of California, Berkeley. Interested 
astronomers elsewhere are also notified immediately.  A Web page on LOSS is at
http://astro.berkeley.edu/$\sim$bait/kait.html.

LOSS found its first supernova
in 1997 --- SN 1997bs; ironically, it might not even be a genuine SN (Van Dyk et al.
2000). In 1998, mostly during the second half of the year, LOSS discovered 20
SNe, thereby breaking the previous single-year record of 15 held by the Beijing
Astronomical Observatory Supernova Search.  In 1999, LOSS doubled this with 40
SNe. In 2000, LOSS found 38 SNe, even though we spent a significant fraction of
the observing time expanding the database of monitored galaxies rather than
searching for SNe. With this expanded database, LOSS discovered 68 SNe in 2001,
82 in 2002, 95 in 2003, and 83 in 2004. We discovered SN 2000A
and SN 2001A, and hence the first supernova of the new millennium, regardless
of one's definition of the turn of the millennium! During the past few years,
KAIT has discovered {\it well over half} of all nearby SNe reported world-wide,
from all searches combined. Thus, KAIT/LOSS is currently the world's most
productive search engine for nearby SNe.

    At the Lick and Keck Observatories, we spectroscopically confirm and classify
nearly all of the SNe that other observers haven't already classified. Thus,
the sample suffers from fewer biases than most.  Already, our observations and
Monte-Carlo simulations have shown that the rate of spectroscopically peculiar
SNe~Ia is considerably larger than had previously been thought (Li et
al. 2001a).

    Follow-up observations for the discovered SNe are emphasized during the
course of LOSS. Our goal is to build up a multicolor database for nearby
SNe. Because of the early discoveries of most LOSS SNe, our light curves
usually have good coverage from pre-maximum brightening to post-maximum
decline. Moreover, LOSS SNe are automatically monitored in unfiltered mode as a
byproduct of our search; these can sometimes be useful for other studies
(e.g., Matheson et al. 2001). The positions of SNe in KAIT images were used to
identify the same SNe at very late times in {\it HST} images (Li et al. 2002), 
allowing us to determine the late-time decline rates.

   LOSS also discovers novae in nearby galaxies (e.g., M31), cataclysmic
variable stars, and occasionally comets (Li 1998; Li et al. 1999). Although it records
many asteroids, we do not conduct follow-up observations of them, so most are 
subsequently lost.

\section{New Morphological Classifications}

 In Papers I and II, morphological classifications were given for the host
galaxies of 177 and 231 SNe, respectively. In Table 1 of the present paper we
list, for an additional 212 SNe, (1) the SN name, (2) the host-galaxy name, (3) the 
SN classification, (4) the type of the host galaxy on the Yerkes system (Morgan 1958, 
1959), (5) the host-galaxy type on the David Dunlap Observatory (DDO) system (van 
den Bergh 1960a, 1960b, 1960c), and (6) the published radial velocity of the SN
host galaxy.  The database examined in the present investigation extends
through the end of the year 2004. 

However, recent careful inspection of the monitoring data of all the host
galaxies classified in Papers I and II reveals that for 15 galaxies, 
the corresponding SNe (discovered and reported by other observers) were 
actually {\it not} successfully imaged by KAIT: 
either the SNe were too faint, or all the KAIT images for a particular
galaxy were plagued by bad weather. Moreover, the host galaxy of SN 1998dl
(NGC 1084) was included in both Papers I and II. We thus need to 
exclude classification for 16 galaxies in our sample, leaving the 
total number of host galaxies classified in Papers I through III 
to be 604.  The 16 galaxies that need to be removed from the study
are listed in Table 2. 

    The Yerkes classification system provides a one-dimensional classification
along the sequence ``a - af - f - fg - g - gk - k." Objects of type ``a" have
the lowest central concentration of light, and those of type ``k" exhibit the
strongest central concentration. In contrast, the DDO system of morphological
classification is three-dimensional. The first DDO classification parameter is
the Hubble type (Hubble 1936), and the second is bar strength measured along
the four-stage sequence S -- S(B?) -- S(B) -- SB.  As a third parameter, the DDO
system uses both spiral-arm morphology and surface brightness to assign
galaxies to luminosity classes I (supergiant), II (bright giant), III (giant),
IV (subgiant), and V (dwarf). In Table 1 uncertain values are followed by a
colon (:), and very uncertain ones by a question mark (?).

     The original Hubble classification system, and its subsequent evolution in
the hands of Sandage (1961), was optimized for the classification of galaxy
images on photographic plates obtained with large reflecting telescopes. On the
other hand the DDO system was devised to classify the lower-resolution images
of galaxies on the Palomar Observatory Sky Survey (POSS). The DDO system is
therefore particularly well suited to the classification of lower-resolution
paper prints of the galaxy images from the POSS-I blue and POSS-I red
surveys. For some galaxies it was also possible to consult the
higher-resolution POSS-II blue images. Furthermore, the KAIT images provide
useful information on the structure of the cores of many images that were
burned out on the POSS. The accuracy and long-term stability of the DDO system
have been discussed in detail in our Paper II. A drawback of the lower-quality
images that can be used for classifications on the DDO system is that they do
not (except in the case of some edge-on galaxies) allow one to distinguish
between elliptical (E) and lenticular (S0) galaxies.

\section{SUPERNOVA CLASSIFICATIONS}

  The spectral classifications of SN type (see Filippenko 1997 for a review)
that are given in Table 1 were drawn from the IAU Circulars. Supernovae of Type
Ia were divided into ``normal" and ``peculiar" categories on the basis of
careful inspection of the spectroscopic information in the IAU
Circulars. Objects that showed the strong Si~II $\lambda$5970 feature or Ti~II
absorption lines near 4200~\AA\ (which are evidence for a subluminous SN
1991bg-like event; Filippenko et al. 1991b), or weak Si~II $\lambda$6150
absorption or strong Fe~III absorption (which indicates a possibly
overluminous, SN 1991T-like event; Filippenko et al. 1991a) were classified as
``peculiar" SNe~Ia. Also in this category are true mavericks such as SN 2000cx
(Li et al. 2001b) and SN 2002cx (Li et al. 2003; not in the LOSS sample), which
cannot be put into the conventional SN~Ia classification scheme.

Out of the 604 SNe that have their host galaxies classified in Papers 
I through III, only 15 SNe (2.5\% of the total) were not spectroscopically
classified. 

\section{DISCUSSION}

\subsection{Frequency Distribution over Hubble Types}

    In Table 3 the combined data from Table 1 of the present paper
and those given in Papers I and II have been sorted by host-galaxy
Hubble type and by supernova type. Galaxies that could not be confidently 
assigned to a Hubble type are excluded. Also, the 16 galaxies listed in Table
2 have been removed from the statistics. In doing the statistics that are 
discussed below, galaxies of intermediate morphology such as Sc/Ir were 
counted as 0.5 Sc and 0.5 Ir.  By the same token, one supernova (SN 2002bt) 
that occurred in UGC 8584, a triple-galaxy system with DDO type
``St + E + S," was counted as 0.33 E, 0.33 St, and 0.33 S. 
The new data show patterns that are broadly similar
to those previously found in Papers I and II.

    A Kolmogorov-Smirnov (K-S) test shows no significant difference
between the distributions of the small numbers of SNe~IIb
and SNe~IIn over Hubble type. Similarly, no significant
difference is found between the distribution over Hubble
types of normal SNe~II and of the combined data for SNe~IIb
and SNe~IIn. In the subsequent discussion the data on all
209 SNe~II have therefore been combined.

     A comparison between the distributions over Hubble types
of normal SNe~Ia and of SNe~II is shown in Figure 1. 
Normal SNe~Ia are common among early-type (E--E/Sa)
galaxies, whereas all types of SNe~II are rare in such early-type
galaxies. A K-S test shows that there is only a
0.01\% probability that the SNe~Ia and SNe~II in our sample were
drawn from the same parent population of morphological types. 

   In Paper II we discussed five SNe~Ibc and SN~II that unexpectedly occurred
in early-type galaxies. Two additional objects of this type occur in the new
data contained in Table 1: SN 2004V, to whose host galaxy we assign type E:0,
and SN 2004X, which occurred in a host that was assigned to type E3.  The host
galaxy of SN 2004V is small ($0\arcmin3\times0\arcmin2$), and our
classification based on the low-resolution images is quite uncertain.  Clearly
it would be important to use images obtained with larger telescopes (or with
{\it HST}) to search for a sub-population of massive young stars in these two
host galaxies that appear to be of very early type. Another approach is to
measure the integrated colors for all the early-type galaxies in our sample,
and search for possible differences between the galaxies with recorded
core-collapse SNe and all the others. This is beyond the scope of the current
paper. However, here we give two examples for which we have some relevant
information. From de Vaucouleurs et al. (1991), we find that NGC 3720, an ``E1"
galaxy that is the host of the Type II SN 2002at, has quite blue colors of $B -
V$ = 0.69 $\pm$ 0.01 mag and $U - B$ = 0.01 $\pm$ 0.03 mag. This suggests that
it does indeed contain a significant young-population component. On the other
hand NGC 2768, an ``E3/Sa" galaxy that is the host of the Type Ib/c SN 2000ds,
has quite red integrated colors of $B - V$ = 0.99 $\pm$ 0.01 mag and $U - B$ =
0.53 $\pm$ 0.01 mag, implying that it is dominated by an old population.

     Inspection of the numbers in Table 3 also shows that most peculiar
SN 1991bg-like SNe~Ia occur in early-type (E or E/Sa) galaxies. On
the other hand the majority of peculiar SN 1991T-like SNe Ia  were discovered
in intermediate-type spirals. Figure 2 shows the Hubble-type distribution of 
the host galaxies of various subclasses of SNe~Ia, and we clearly
see the dichotomy between early-type hosts for the SN 1991bg-like objects
and late-type hosts for the SN 1991T-like ones. A K-S test shows that
there is only a 0.1\% probability that the SN 1991T-like and the SN 
1991bg-like objects were drawn from the same parent
population. The observed difference is in the sense that would be
expected if the more luminous SN 1991T-like objects have younger
progenitors than do the fainter SN 1991bg-like objects. A K-S test
shows that the distribution over Hubble type of the 12 SN
1991T-like SNe Ia does not differ significantly from
that of ``normal" SNe~Ia. On the other hand there is only a 0.01\%
probability that the normal SNe~Ia and the SN 1991bg-like ones  were drawn
from the same parent population. The observed difference is in
the sense that would be expected if the subluminous SN 1991bg-like
SNe Ia (which mostly occur in E and E/Sa galaxies) typically have
old progenitors. Similar results have previously been obtained by
Hamuy et al. (1996, 2000) and by Howell (2001).

          A comparison between the distributions over Hubble types of
normal SNe~Ia and SNe~Ibc shows that there is only a 0.04\%
probability that these two samples were drawn from the same parent
population. On the other hand a K-S test shows no significant
difference between the distributions over Hubble types of SNe Ibc
and the sum of all three subtypes of SNe~II. It is therefore concluded
that SNe Ibc and SNe II occur among similar stellar populations.

         It should be noted that the frequency distributions
discussed above may be affected by several selection effects and
observational biases. For example, the distribution reflects the
SNe discovered in the sample of galaxies monitored by LOSS. As
discussed by Li et al. (2001a), the LOSS sample galaxies were 
selected from several large galaxy catalogs, and the very late-type 
spiral (Scd, Sd, and Sdm) and irregular (Ir) galaxies are 
underrepresented. More generally, galaxies having low optical 
luminosity or low surface brightness are underrepresented. 
Observational biases, such as the Malmquist
bias caused by the differences in the intrinsic luminosities of SNe,
may also affect the apparent frequency distribution of the 
host-galaxy types. A more detailed discussion of the various observational
biases that affect the discovery rate of SNe~Ia can be found in Li,
Filippenko, \& Riess (2001). The intrinsic frequency distributions
of various types of SNe in galaxies of different Hubble types (i.e.,
the SN rates) will need to consider all of the selection and
observational biases. The SN rate calculation for LOSS is
currently being investigated, and the initial results are reported
by Leaman, Li, \& Filippenko (2004). Finally, inspection of the data
in Table 3 suggests that one of us (S.vdB.) had a strong classification bias
in favor of Hubble types Sa, Sb, and Sc, and against the intermediate
types Sab and Sbc.

\subsection{Frequency Distribution over Broader Morphological Classes}

         The images of many of the distant host galaxies are so small
that it is not possible to assign them with confidence to a Hubble
type. Nevertheless, many of these objects can be placed in the
broader ``spiral" category. Furthermore, it is often difficult (or
impossible) to distinguish between E and S0 galaxies on the Schmidt images
of the Palomar Sky Survey. Consequently, only highly flattened
[$(1-b/a) \approx 0.7$] objects are classified as being of type S0 on the
DDO system. In order to take maximum advantage of the present
observational material we have therefore sorted the supernova host
galaxies into morphological classes E, S0, S, Ir, other, and ``?"
(Table 4). Again, galaxies of intermediate morphology were counted 
in all possible morphologies according to their probabilities.
SN 1999gf, for example, with a host galaxy having a DDO type of 
``cD or E/Sa,"  was counted as 0.25 E and 0.25 S in Table 4. 

        These data allow one to compare the distribution of
200 normal SNe~Ia with that of 251 SNe of types II, IIb, and IIn.
A K-S test shows that there is only a 0.3\%
probability that these two samples were drawn from the same parent
population. This result is less significant than the 0.01\%
probability that was previously found from the data in Table 2,
showing that the confidence in our results is more
dependent on accurate morphological classifications than on
sample size. A similar conclusion may be drawn from a comparison
of the 200 normal SNe Ia and the 88 SNe~Ibc in Table 4. A K-S test
shows that the probability that these samples were drawn from the
same parent population is 1\%, compared to a 0.04\% probability found
from the smaller number of normal SNe~Ia and the SNe~Ibc in Table 3.
Clearly, fine morphological subdivision is important when the
properties of supernovae are a sensitive function of the Hubble
types of their host galaxies.

\subsection{Frequency Distribution over Yerkes Morphological Classes}

      In the Yerkes classification system (Morgan 1958, 1959) galaxies
are classified according to their central concentration of light.
Such a classification system has the advantage that it is more
easily adapted to automatic digital classification than is
Hubble's tuning-fork system. Yerkes classifications of the host
galaxies of newly discovered supernovae are listed in Table 1.
As expected, these data show that the host galaxies of normal
SNe~Ia are, on average, more centrally concentrated than are
those of SNe~II (including SNe~IIb and IIn). However, mainly due to 
the smaller database of Yerkes types, this result is of lower
statistical significance than the comparable result from the
Hubble types of host galaxies that was reported in \S~4.1.
The Yerkes classifications also confirm (albeit at a lower level
of statistical confidence than from the larger sample of host 
galaxies having Hubble classifications) that the distribution of
concentration classes of SNe~II and of SNe~Ibc are
indistinguishable. Finally, almost all SN 1991bg-like SNe~Ia 
are found to have occurred in compact host galaxies
of Yerkes class ``k."

\section{Conclusions}

       A uniform sample of 604 host galaxies of recent supernovae has
been classified on the DDO system. These data lead to the following
conclusions.

\begin{enumerate}

\item{The distributions of the morphological types of the host galaxies
      of SNe~Ia and SNe~II differ at a very high level of statistical
      significance, with SNe~Ia favoring earlier-type galaxies.}

\item{The distribution of the morphological types of host galaxies of
      SNe~Ibc is indistinguishable from that of SNe~II.}

\item{The distribution over morphological types of small numbers of
      SNe~IIn and SNe~IIb do not appear to differ from that of normal
      SNe~II.}

\item{SN 1991T-like SNe~Ia occur mainly in host galaxies of intermediate
      morphological types, whereas SN 1991bg-like SNe~Ia are mostly seen in E and
      E/Sa galaxies. This observed difference is significant at the 99.9\%
      level.}

\item{A few SNe~II were detected in E galaxies. Higher-resolution 
      images will be required to establish if some of these
      galaxies were misclassified, or if they might be old galaxies in
      which a small young-population component is embedded. Comparison of 
      the integrated colors may also help to reveal possible differences
      between the early-type galaxies with recorded core-collapse 
      SNe and those without.}

\end{enumerate}

\acknowledgments

The work of A.V.F.'s group at U. C. Berkeley is supported by National Science
Foundation grant AST-0307894, as well as by NASA grant GO-10182 from the Space
Telescope Science Institute, which is operated by AURA, Inc., under NASA
contract NAS 5-26555.  KAIT was made possible by generous donations from Sun
Microsystems, Inc., the Hewlett-Packard Company, AutoScope Corporation, Lick
Observatory, the National Science Foundation, the University of California, and
the Sylvia and Jim Katzman Foundation. A.V.F. is grateful for a Miller Research
Professorship at U.C.  Berkeley, during which part of this work was
completed. S.vdB. thanks Jasper Wall for his advice.

\begin{figure}
\psfig{figure=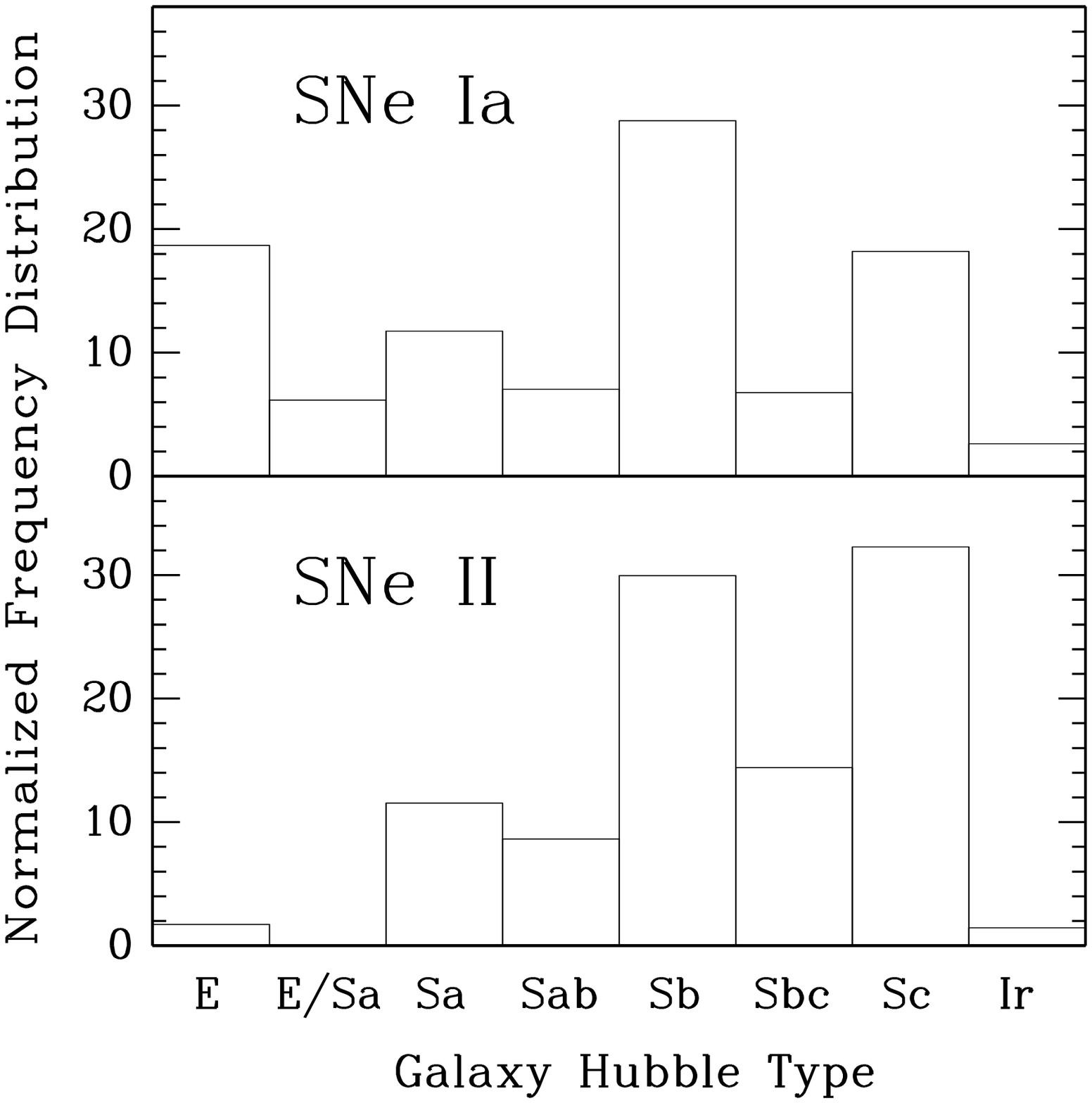,height=5.0truein,width=7.2truein,angle=-90} 
\caption{Normalized (total = 100) frequency distribution of SNe~Ia and SNe~II 
versus host-galaxy Hubble types.}
\end{figure}

\begin{figure}
\psfig{figure=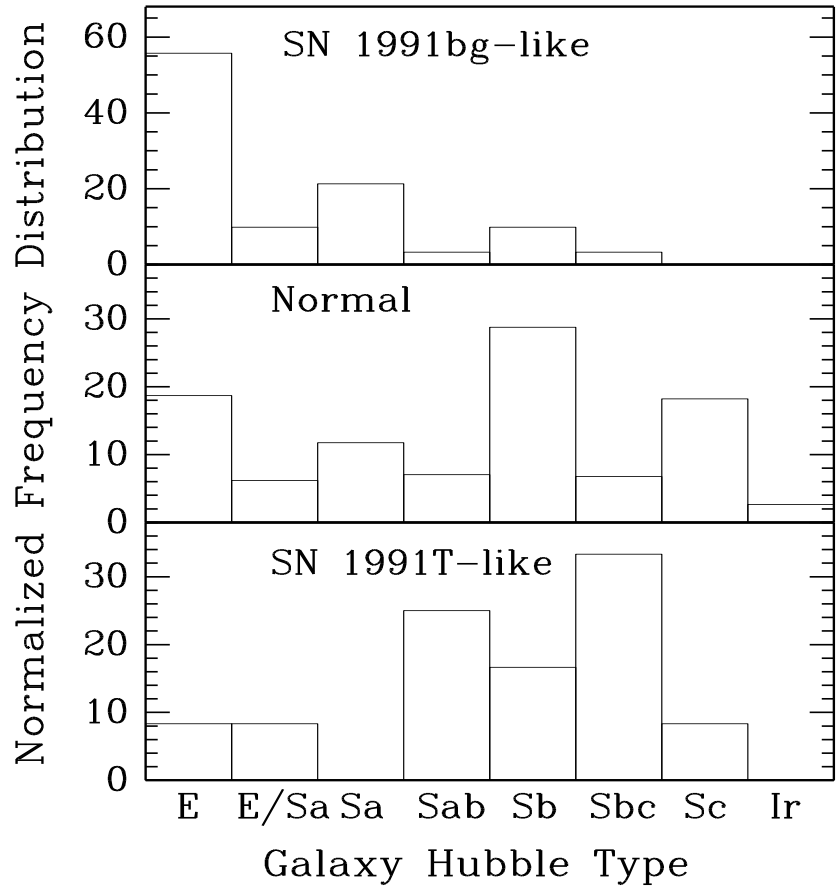,height=5.0truein,width=7.2truein,angle=-90} 
\caption{Normalized (total = 100) frequency distribution of SN 1991T-like,
normal, and SN 1991bg-like SNe~Ia 
versus host-galaxy Hubble types.}
\end{figure}

\renewcommand{\arraystretch}{0.78}

\begin{deluxetable}{lllcllc}
\tablecaption{Classifications of SN Host Galaxies}
\label{1}
\tablehead{
\colhead{SN} & \colhead{Galaxy} & \colhead{SN} &\colhead{Yerkes}  & \colhead{DDO} & \colhead{Redshift} &  \colhead{Remarks}\\
             &                  & \colhead{Type} & \colhead{Type} & \colhead{Type} & \colhead{(km/s)} & 
}
\startdata 
1998C   & U3825         & II &    fg & Sbc II      &  8281  &      \\
1998S   & N3877         & IIn&     f & Sbc         &   895  &  (2) \\
1998aq  & N3982         & Ia &    fg & Sc: II:     &  1109  &      \\
2000dx  & U1775         & Ia &    fg & Sc pec      &  9108  &  (3) \\
2000ej  & I1371         & Ia-pec (91bg) & k & E2   &  9102  &      \\
2000fe  & U4870         & II &    gk & Sb          &  4218  &      \\
2000fm  & N1612         & II &    fg & Sbc         &  \nodata & \\
2000fo  & P70148        & Ia &     g & Sab         &  7152  &  \\
2001U   & N5442         & Ia &     g & Sb: t       &  8517  &  \\
2001ah  & U6211         & Ia-pec (91T)&f&Sbc I     & 16788  &  \\
2001ak  & U11188        & II &    af:& Sc:         &  5285  &  (2) \\
2001bb  & I4319         & Ic &     g & Sab:        &  4653  &  \\
2001gb  & I582          & Ia &     g & S           &  7714  &  \\
2001gc  & U3375         & Ia &    fg & Sc II:      &  5783  &  \\
2001hf  & M-03-23-17    & II &     ? & S           &  4486  & (12) \\
2001hh  & M-02-57-22    & II &    gk & Sa          &  7445  &  \\
2002ct  & Anonymous     & unknown &gk& S0:         &  \nodata & (2) \\
2002fk  & N1309         & Ia &    fg & Sc          &  2136  &    \\
2002kg  & N2403         & IIn&    fg & Sc III      &   131  &    \\
2003bt  & M-01-28-06    & Ia &    fg & S(B?)c      &  7972  &    \\
2003cb  & N4885         & II &    gk & Sa?         &  3366  &   \\
2003db  & M+05-23-21    & II &     g & Sab:        &  8067  &   \\
2003eg  & N4727         & II &     g & S(B)b II:   &  7495  &      \\
2003eh  & M+01-29-03    & Ia &    fg & Sb pec      &  7651  &      \\
2003ei  & U10402        & IIn &    ? & St + Pec    &\nodata &  (5)  \\
2003ej  & U7820         & IIb &    f & Sc II       &  5090  &      \\
2003ek  & anonymous     & Ia-pec (91bg) &     f:& S           & 10804  &  (2)  \\
2003el  & N5000         & Ic &    fg & SBbc I      &  5608  &      \\
2003em  & ESO 478-G6    & Ia &    fg & Sc I        &  5332  &      \\
2003ep  & N7053         & Ia &     k & E2/Sa       &  4708  &      \\
2003ev  & anonymous     & Ic &     g & Sab         &  7200  &      \\
2003ez  & PGC 42782     & Ia &     g & Sb pec      & 14343  &      \\
2003fa  & M+07-36-33    & Ia-pec (91T) &     g & Sb: t       &  1800  &      \\
2003fb  & U11522        & II &     g & Sc:         &  5259  &      \\
2003fc  & M-03-51-05    & Ic &    fg & S           & 10400  &      \\
2003fd  & U8670         & Ia &    fg & Sc: II:     & 17911  &      \\
2003gd  & N628          & II &    fg & Sc I        &   657  &      \\
2003gf  & M-04-52-26    & Ic &     ? & Pec         &  2600  &      \\
2003gg  & I1321         & II &     g & S(B?)b II   &  6660  &      \\
2003gi  & I1561         & Ia &     f & Sbc         &  3899  &      \\
2003gj  & N7017         & Ia-pec (91bg) &     k:& E1 + E0     & 10119  &  (1)  \\
2003gk  & N7460         & Ib &     g & Sc II:      &  3192  &      \\
2003gl  & N7782         & Ia &     g & Sb II       &  5379  &      \\
\enddata
\tablenum{1}
\end{deluxetable}

\begin{deluxetable}{lllcllc}
\tablecaption{(continued)}
\label{1}
\tablehead{
\colhead{SN} & \colhead{Galaxy} & \colhead{SN} &\colhead{Yerkes}  & \colhead{DDO} & \colhead{Redshift} &  \colhead{Remarks}\\
             &                  & \colhead{Type} & \colhead{Type} & \colhead{Type} & \colhead{(km/s)} & 
}
\startdata 
2003gm  & N5334         & IIn &    f & S(B?) III-IV&  1382  &      \\
2003gn  & CGCG 452-024  & Ia &    gk & Sab         & 10328  &      \\
2003go  & ESO 595-G001  & IIn &    g & Sa:         & 10765  &      \\
2003gp  & U10160        & II &    gk & SBab        &  9967  &      \\
2003gq  & N7407         & Ia-pec (91T) & f & Sbc II      &  6430  &      \\
2003gr  & M-04-55-14    & Ia &     g:& SBb         &  7691  &      \\
2003gs  & N936          & Ia-pec (91bg)& k & SBa         &  1430  &      \\
2003gt  & N6930         & Ia &     g & Sb t        &  4694  &      \\
2003gu  & U12331        & IIb &    g & Sab:        &  5794  &       \\
2003gv  & M+05-03-66    & II &    fg & Sbc:        & 10423  &       \\
2003gw  & U3252         & II &     f & Sc I-II     &  6115  &       \\
2003hc  & U1993         & II &     ? & S           &  8018  &  (2)   \\
2003hd  & M-04-05-10    & II &     g & Sbc:        & 11842  &       \\
2003he  & M-01-01-10    & Ia &    fg & Sb:         &  7649  &       \\
2003hf  & U10586        & II &     g & Sab         &  9384  &       \\
2003hg  & N7771         & II &     f & S pec       &  4277  &  (3)?  \\
2003hh  & U12890        & Ia-pec (91bg) &     k & E4          & 11602  &       \\
2003hk  & N1085         & II &     g & Sb II       &  6789  &       \\
2003hl  & N772          & II &     f & Sbc t?      &  2472  &       \\
2003hm  & U2295         & Ia &     g & Sb          &  4172  &       \\
2003hp  & U10942        & Ic-pec &fg & Sb t?       &  6378  &       \\
2003hs  & U11149        & Ia-pec (91bg) & k & E3/Sa       & 14990  &       \\
2003ht  & U2457         & II &     g & Sab:        & 10218  &       \\
2003hv  & N1201         & Ia &     k & E4          &  1671  &       \\
2003hw  & anonymous     & Ia &     g & Sb          &\nodata &  (6)   \\
2003hx  & N2076         & Ia &    gk & Sa          &  2142  &  (4)   \\
2003hy  & I5145         & IIn &    g & Sb II       &  7355  &       \\
2003hz  & PGC 17866     & Ia &    fg & Sb:         &  6047  &       \\
2003ib  & M-04-48-15    & II &     g & Sb          &\nodata &       \\
2003ic  & M-02-02-86    & Ia &     g & E2/Sa       & 16690  &       \\
2003id  & N895          & Ic &    fg & Sbc I-II    &  2288  &       \\
2003ie  & N4051         & II &     f & Sc II       &   700  &       \\
2003if  & N1302         & Ia &     k & Sa          &  1703  &       \\
2003ig  & U2971         & Ic &    fg & Sb:         &  5881  &       \\
2003ih  & U2836         & Ibc &    k & E:1         &  4963  &       \\
2003ik  & U4185         & Ia &     ? & Sc          &  7115  &  (2)   \\
2003im  & anonymous     & Ia &     k & Sa          &  5804  &       \\
2003in  & I1956         & Ia &    fg:& Sb          &  6401  &       \\
2003ip  & U327          & II &     g & Sa          &  5398  &       \\
2003iq  & N772          & II &    fg & Sbc t?      &  2472  &       \\
2003ir  & U3726         & II &     g & Sb          &  7657  &       \\
2003is  & U11430        & Ic &     f & Sc          &  5482  &       \\
2003it  & U40           & Ia &     g & SBb         &  7531  &       \\
\enddata
\tablenum{1}
\end{deluxetable}

\begin{deluxetable}{lllcllc}
\tablecaption{(continued)}
\label{1}
\tablehead{
\colhead{SN} & \colhead{Galaxy} & \colhead{SN} &\colhead{Yerkes}  & \colhead{DDO} & \colhead{Redshift} &  \colhead{Remarks}\\
             &                  & \colhead{Type} & \colhead{Type} & \colhead{Type} & \colhead{(km/s)} & 
}
\startdata 
2003iv  & M+02-08-14    & Ia &     k & E1          & 10285  &       \\
2003iw  & N7102         & II &     f & Sc:         &  4866  &       \\
2003ix  & U3746         & Ia &     g & Sa          &  7668  &       \\
2003iz  & U638          & Ia-pec (91bg) &     k & E0          & 14453  &       \\
2003ja  & N846          & II &    fg & SBb II      &  5118  &      \\
2003jc  & M-01-58-18    & II &     f & Sc:         &  6029  &      \\
2003jd  & M-01-59-21    & Ic-pec & f & Sb:         &  5654  &      \\
2003je  & N2668         & II &    gk & S(B)bc      &  7529  &      \\
2003jh  & M-02-11-30    & IIn &   fg & Sbc II      &  8898  &      \\
2003jz  & U5225         & Ia &     k & E0          &  4906  &      \\
2003ka  & M+06-50-20    & II &     f & Sc III-IV   &  5761  &      \\
2003kb  & U3432         & Ic &     ? & S0/Sb       &  4998  &  (2)  \\
2003kc  & M+05-23-37    & Ia &    fg & Sc I?       & 10003  &      \\
2003kd  & U2468         & Ia &     k & E0/Sa       &\nodata &      \\
2003ke  & M+06-22-09    & IIn &   fg & Sb:         &  6176  &      \\
2003kf  & M-02-16-02    & Ia &    af & S IV        &  2215  &      \\
2003kw  & M+05-27-49    & II &     g & Sa (pec?)   &  8012  &      \\
2003ld  & U148          & II &     ? & Sc?         &  4213  &  (2)  \\
2003lo  & N1376         & II &    fg & Sc I        &  4155  &      \\
2003lp  & U6711         & II &    gk & Sa          &  2702  &      \\
2003lq  & U5            & Ia &    fg & Sb II       &  7271  &      \\
2003ls  & PGC 11402     & Ia &    fg & ?           & 13000  &      \\
2004A   & N6207        &  II &     f & Sc/Ir       &   852  &      \\
2004C   & N3683        &  Ic &     f & S           &  1716  &  (2)  \\
2004D   & U6916        &  II &    fg & Sb          &  6182  &      \\
2004E   & PGC 46239    &  Ia &    gk & Sa pec      &  8936  &      \\
2004F   & N1285        &  IIn &    g & Sc          &  5239  &      \\
2004G   & N5668        &  II &     f & Sc III-IV   &  1583  &      \\
2004H   & I708         &  Ia-pec (91bg) &     k & E2          &  9497  &  (7)  \\
2004I   & N1072        &  II &     g & Sb          &  8018  &      \\
2004J   & ESO 554-G33  &  Ia &    fg & S           &\nodata &      \\
2004K   & ESO 579-G22  &  Ia &    gk:& S(B)b:      & 10832  &      \\
2004L   & M+03-27-38   &  Ia &     g & S(B)b       &  9686  &      \\
2004P   & U8561        &  Ia &    fg & Sc          &  7120  &      \\
2004Q   & ESO 507-G11  &  II &     ? & Sc pec?     &  7483  &      \\
2004T   & U6038        &  II &    gk & Sa          &  6437  &      \\
2004U   & anonymous    &  II &    gk:& SBb         &\nodata &      \\
2004V   & anonymous    &  II &     k:& E:0         & 12500  &      \\
2004W   & N4649        &  Ia-pec (91bg) & ? & E1          &  1117  &  (8)  \\
2004X   & anonymous    &  II &     k & E3          &  3917  &      \\
2004Y   & anonymous    &  Ia &     k & E2          & 20760  &      \\
2004ab  & N5054         & Ia &    fg & Sc I        &  1741  &      \\
2004ak  & U4436         & II &     f?& S           &  7214  &  (2)  \\
\enddata
\tablenum{1}
\end{deluxetable}

\begin{deluxetable}{lllcllc}
\tablecaption{(continued)}
\label{1}
\tablehead{
\colhead{SN} & \colhead{Galaxy} & \colhead{SN} &\colhead{Yerkes}  & \colhead{DDO} & \colhead{Redshift} &  \colhead{Remarks}\\
             &                  & \colhead{Type} & \colhead{Type} & \colhead{Type} & \colhead{(km/s)} & 
}
\startdata 
2004al  & ESO 565-G25   & II &     g & Sa          &\nodata &      \\
2004am  & N3034         & II &     ? & Pec         &   203  &      \\
2004an  & I4483         & II &    fg & Sa          &  8979  &      \\
2004ao  & U10862        & Ib &     f & SBb         &  1691  &      \\
2004ap  & PGC 29306     & Ia &     k & E2          &  7177  &      \\
2004aq  & N4012         & II &     g:& Sa          &  4182  &      \\
2004as  & anonymous     & Ia &    af & S/Ir        &  9300  &  (9)  \\
2004at  & M+10-16-37    & Ia &     ? & Ir ?        &  6935  &      \\
2004au  & M+04-42-2     & II &     g & Sa          &  7800  &      \\
2004av  & ESO 571-G15   & Ia &     ? & S           &  7057  &  (2)  \\
2004aw  & N3997         & Ic &     ? & St + St     &  4771  &  (1)  \\
2004ax  & N5939         & Ibc &    g & Sbc         &  6687  &      \\
2004ay  & U11255        & IIn &    ? & Sc/Ir       &  9723  &  (2)  \\
2004az  & U6853         & Ia &     k & E:4         &  8639  &      \\
2004bd  & N3786         & Ia &     g & Sb pec      &  2678  &  (3)  \\
2004be  & ESO 499-G34   & II &    af & S IV:       &  2282  &      \\
2004bf  & U8739         & Ic &     ? & S           &  5032  &  (2)  \\
2004bh  & U5161         & II &     g & S/Ir        & 10079  &      \\
2004bi  & U5894         & IIb &    g & Sb          &  6537  &      \\
2004bj  & M+01-34-13    & Ia &     k & E0          & 15033  &      \\
2004bk  & N5246         & Ia &    gk & SBb         &  6906  &      \\
2004bl  & M+00-31-42    & Ia &     ? & S/Ir        &  5192  &  (2)  \\
2004bm  & N3437         & Ic &     g:& Sbc ?       &  1283  &      \\
2004bn  & N3441         & II &     g & Sa:         &  6533  &      \\
2004bo  & ESO 576-G54   & Ia &     k & E3          &  7024  &      \\
2004bq  & ESO 597-G32   & Ia &    gk & Sa:         &\nodata &      \\
2004br  & N4493         & Ia-pec (91T/00cx) &     k & E1 t?       &  6943  &      \\
2004bs  & N3323         & Ib &    fg & S(B?)b      &  5164  &      \\
2004bt  & U9178         & unknown  &  f    & S(B?)c:     &  8704  &      \\
2004bv  & N6907         & Ia-pec (91T) &     ? & S pec       &  3161  &  (3)  \\
2004bw  & M+00-38-19    & Ia &    fg & Sc          &  6355  &      \\
2004by  & N7116         & II-pec & ? & Sb          &  3532  &  (3)? \\
2004bz  & M+02-56-25    & Ia &     g & Sab:        & 10232  &      \\
2004ca  & U11799        & Ia &     ? & S           &  5338  &  (10) \\
2004cc  & N4568         & Ic &     f:& S pec       &  2255  &  (3)  \\
2004ci  & N5980         & II &     g & Sb          &  4092  &      \\
2004cm  & N5486         & II &     g & Sbc:        &  1390  &      \\
2004cq  & U9882         & Ia &     ? & S           &  6595  &  (2)  \\
2004cs  & U11001        & Ibc &    f & Sc pec      &  4215  &      \\
2004cu  & N5550         & II &    fg:& Sbc:        &  7427  &      \\
2004db  & N7377         & Ia &     k & E:2         &  3351  &      \\
2004dc  & I1504         & Ic &    fg & Sb:         &  6271  &      \\
2004dd  & N124          & II &    fg & Sc          &  4060  &      \\
\enddata
\tablenum{1}
\end{deluxetable}

\begin{deluxetable}{lllcllc}
\tablecaption{(continued)}
\label{1}
\tablehead{
\colhead{SN} & \colhead{Galaxy} & \colhead{SN} &\colhead{Yerkes}  & \colhead{DDO} & \colhead{Redshift} &  \colhead{Remarks}\\
             &                  & \colhead{Type} & \colhead{Type} & \colhead{Type} & \colhead{(km/s)} & 
}
\startdata 
2004dh  & M+04-01-48    & II &     f & S           &  5794  &      \\
2004dj  & N2403         & II &     ? & Sc III      &   131  &  (11) \\
2004dk  & N6118         & Ic &     f & Sbc II      &  1573  &      \\
2004dn  & U2069         & Ic &     ? & Sc III-IV   &  3779  &      \\
2004dr  & ESO 479-G42   & II &    af & S pec       &  6917  &      \\
2004ds  & N808          & II &     f & Sb II       &  4964  &      \\
2004dt  & N799          & Ia &     g:& Sbc         &  5915  &      \\
2004du  & U11683        & IIn &    ? & S           &  5025  &  (2)  \\
2004dv  & M-01-06-12    & II &     f & Sc pec?     &  4754  &      \\
2004dy  & I5090         & II &     g:& Sb:         &  9340  &  (3)? \\
2004dz  & anonymous     & Ia &     f & S/Ir        &\nodata &      \\
2004ea  & M-03-11-19    & Ia &    af & S pec       &  1953  &      \\
2004eb  & N6387         & II &     ? & St  ?       &  8499  &  (1)  \\
2004ef  & U12158        & Ia &     g:& Sc I        &  9290  &      \\
2004eg  & U3053         & II &     ? & Sc ?        &  2407  &      \\
2004ep  & I2152         & II &    gk & Sb II:      &  1875  &      \\
2004er  & M-01-07-24    & II &    fg & Sbc:        &  4411  &      \\
2004es  & U3825         & II &    fg & Sc:         &  8281  &      \\
2004et  & N6946         & II &     f & Sc I        &    48  &      \\
2004ex  & N182          & II &    gk & Sb          &  5261  &      \\
2004ez  & N3430         & II &     g & Sc II       &  1586  &      \\
2004fc  & N701          & II &     g & S pec       &  1829  &      \\
2004fe  & N132          & Ic &     g:& Sc          &  5361  &      \\
2004ff  & ESO 552-G40   & Ic &     gk& Sb          &  6790  &      \\
2004fg  & M+05-56-07    & Ia &     fg& Sc          &  9034  &      \\
2004fx  & M-02-14-03    & II &     ? & S           &  2673  &  (2)  \\
2004gd  & N2341         & IIn &    gk& Sab:        &  5227  &      \\
2004ge  & U3555         & Ic &     g & Sc t?       &  4835  &      \\
2004gg  & U5234         & II &     f & Sc:         &  6017  &      \\
2004gh  & M-04-25-06    & II &     g & S(B?)b      &  3662  &      \\
2004gi  & M-05-25-32    & Ia &     f & Sc          &  3244  &      \\
2004gj  & I701          & IIb &    f & Sc          &  6143  &      \\
2004gk  & I3311         & Ic &     ? & S           &  -122  &  (2)  \\
2004gm  & M-02-33-80    & Ia &     f & Sab         &  4975  &      \\
2004gn  & N4527         & Ic &    fg & Sbc         &  1736  &      \\
2004go  & I270          & Ia &   k   & E1          &  8745  &      \\
2004gq  & N1832         & Ic &   g   & S(B?)bc II  &  1939  &      \\
2004gr  & N3678         & II &   g   & Sc:         &  7210  &      \\
2004gs  & M+03-22-20    & Ia-pec (91bg) & gk & Sa  &  7988  &      \\
2004gt  & N4038         & Ic &   a:  & Sc? pec t   &  1642  &  (3) \\
\enddata
\tablenotetext{}{Note: (1) merger; (2) edge-on; (3) tides; (4) dusty; (5) SN closest
to peculiar galaxy; (6) SN in small distant galaxy, not in nearer large SBb;
(7) might also be classified E2/Sa; (8) our images of M60 (= NGC 4649) are
overexposed, so the adopted E1 classification is from van den Bergh (1960c);
(9) has bright Sc~II companion; (10) strong Galactic foreground absorption
possible; (11) galaxy too large to classify with present images, so we have
adopted the Sc~III classification from van den Bergh (1960c); and 
(12) bright foreground star superimposed on the nucleus.}
\end{deluxetable}

\begin{deluxetable}{llccll}
\tablecaption{SNe and Host Galaxies (Papers I, II) Now Excluded from the Sample}
\label{2}
\tablehead{
\colhead{SN} & \colhead{Galaxy} & \colhead{SN Type} &\colhead{Yerkes Type} &\colhead{DDO Type}  & \colhead{Redshift}} 
\startdata 
1998dl &   N1084           &   II   & f    &  Sc II:    &    1406  \\
1998ey &   N7080           & Ic-pec & fg   &  S(B)bc I  &    4839 \\
2000ce &   U4195           &   Ia   & fg   &  SBb II    &    4888 \\
2000cr &   N5395           &   Ic   & fg:  &  Sbt I?    &    3491 \\
2001bg &   N2608           &   Ia   & fg   &  Sc II:    &    2135 \\
2001dp &   N3953           &   Ia   & g    &  S(B)bc I  &    1052 \\
2001dr &   N4932           &   II   & fg   &  S(B?)c II &    7088     \\
2001eg &   U3885           &   Ia   & g:   &  Sbc III:  &    3809 \\
2002bp &   U6332           & unknown& gk   &  Sa:       &    6227\\
2002cv &   N3190           &   Ia   & g    &  Sab t     &    1271\\
2002ed &   N5468           &   II   & f    &  Sc I-II   &    2845 \\
2002jo &   N5708           &   Ia   & fg:  &  S pec     &    2751 \\
2002cy &   N1762           &unknown & g    &  Sab:      &    4753 \\
2003C &    U439            &   II   & g    &  Sa pec    &    5302 \\
2003U &    N6365           &   Ia   & g    &  Sbn:      &    8496 \\
2003X &    U11151          &   Ia   & fg   &  S0:       &    7017 \\
\enddata
\end{deluxetable}

\begin{deluxetable}{llllllll}
\label{3}
\tablecaption{Host-Galaxy Classification and SN Type\tablenotemark{a,b}}
\tablehead{
\colhead{Galaxy type\tablenotemark{c}} & \colhead{Ia} & \colhead{Ia(T)\tablenotemark{d}} & \colhead{Ia(bg)\tablenotemark{e}} &
\colhead{Ibc\tablenotemark{f}} & \colhead{II} & \colhead{IIb}& \colhead{IIn}
}
\startdata
E     &     31.83\tablenotemark{g} &  1 & 17   &  1 &  3  &  0  &  1 \\
E/Sa  &     10.5  &  1 &  3   &  1 &  0  &  0  &  0 \\
Sa    &     20    &  0 &6.5   &  4 & 20  &  0  &  3 \\
Sab   &     12    &  3 &  1   &  7 & 15  &  1  &  1 \\
Sb    &     49    &  2 &  3   & 15 & 52  &  2  &  6 \\
Sbc   &     11.5  &  4 &  1   & 19 & 25  &  3  &  4 \\
Sc    &     31    &  1 &  0   & 21.5&56  &  3  &  9 \\
Sc/Ir &     0     &  0 &  0   &  0  & 1  &  0  &  1 \\
Ir    &     4.5   &  0 &  0   &  0  & 2.5&  0  & 0.5\\
\hline
Total &  170.33   & 12 & 31.5 & 68.5&174.5 & 9 & 25.5\\
\enddata
\tablenotetext{a}{All host galaxies of SNe discovered during LOSS/LOTOSS --
i.e.,  the sum of the data from the present paper and those from Papers I 
and II.}
\tablenotetext{b} {Half-integer values refer to intermediate 
morphologies (e.g., E/Sb is counted as 0.5 E and 0.5 Sb).}
\tablenotetext{c}{Includes S(B) and SB.}
\tablenotetext{d}{SN 1991T-like SN~Ia.}
\tablenotetext{e}{SN 1991bg-like SN~Ia.}
\tablenotetext{f}{Here, the ``Ibc" designation includes
SNe~Ib, Ic, and Ib/c.} 
\tablenotetext{g}{The fractional number 0.83 comes from 0.50 + 0.33,
due to SN 2002bt which occurred in a triple-galaxy system; see
text for details.}
\end{deluxetable}

\begin{deluxetable}{llllllll}
\label{4}
\tablecaption{Frequency Distribution of Broad Morphological Types}
\tablehead{
\colhead{Galaxy type} & \colhead{Ia} & \colhead{Ia(T)} & \colhead{Ia(bg)} &
\colhead{Ibc} & \colhead{II} & \colhead{IIb}& \colhead{IIn}
}
\startdata
E     &     37.58\tablenotemark{a} &  1.5 & 18.5 & 1.5&  3  &  0  &  1 \\
S0    &     2.5   &  0   &  1.5 &  1 &  2  &  1  &  0 \\
S     &    146.92\tablenotemark{b} & 12.5 & 15   &82.5& 193.5& 11  &27.5\\
Ir    &     4.5   &  0   &  0   &  0 &  3  &  0  &  1 \\
Other &     3.5   &  0   &  0   &  2 &  4  &  0  &  1.5 \\
?     &      5    &  0   &  1   &  1 &  1.5&  0  &  1 \\
\hline
Total &  200      & 14 & 36   & 88  &207   & 12 & 32\\
\enddata
\tablenotetext{a} {The fractional number 0.58 comes from 0.33 + 0.25,
due to SN 2002bt (which occurred in a triple-galaxy system) and
SN 1999gf (with a DDO type of ``cD or E/Sa"); see text for details.}
\tablenotetext{a} {The fractional number 0.92 comes from 0.67 + 0.25,
which is due to SN 2002bt (occurred in a triple galaxy system) and
SN 1999gf (with a DDO type of ``cD or E/Sa."). See text for details.}
\end{deluxetable}
\end{document}